\documentclass[10pt,twocolumn]{IEEEtran}
\usepackage{soul,xcolor}
\usepackage{bm}
\include{graphics}
\include{amsmath}
\usepackage[normalem]{ulem}
\usepackage{multirow,enumitem}
\usepackage{algorithm}
\usepackage{algpseudocode}
\usepackage{verbatim}
\usepackage[latin1]{inputenc}
\usepackage{amsmath}
\usepackage{amsfonts}
\usepackage{amssymb}
\usepackage{mathrsfs}
\usepackage{booktabs}
\usepackage{url}
\usepackage[skip=1pt,font=scriptsize]{subcaption}
\usepackage{ifthen}
\usepackage{xspace}
\usepackage{dsfont}
\usepackage{bbm}
\usepackage{cite}
\usepackage{array}
\usepackage{algpseudocode}
\usepackage{algorithm}
\usepackage{breqn}
\usepackage{esint}
\usepackage[skip=2pt,font=footnotesize]{caption}
\usepackage{amsthm}
\usepackage{setspace}
\usepackage{lipsum}
\usepackage{dblfloatfix}
\theoremstyle{plain}

\theoremstyle{definition}

\theoremstyle{remark}

\usepackage{footnote}

\usepackage{bbold}
\usepackage{todonotes}

\usepackage{stmaryrd}
\usepackage{algorithm,algcompatible}
\algnewcommand\INPUT{\item[\textbf{Input:}]}%
\algnewcommand\OUTPUT{\item[\textbf{Output:}]}%
\usepackage{cite}

\DeclareGraphicsExtensions{.eps}

\newcommand{\round}[1]{\ensuremath{\left\lfloor#1\right\rceil}}
                                                          
\title{Millimeter Wave Beam Recommendation\\ via Tensor Completion}
\author{Tzu-Hsuan~{Chou}, Nicolo {Michelusi}, David J. {Love}, and James V. {Krogmeier}
\thanks{The authors are with the School of Electrical and Computer Engineering, Purdue University, West Lafayette, IN, USA; emails: \{chou59, michelus, djlove, jvk\}@purdue.edu.}%
\thanks{This research has been funded by NSF under grant CNS-1642982.}
\vspace{-5mm}
}

\begin{document}
\maketitle
\thispagestyle{empty}
\pagestyle{empty}
\setulcolor{red}
\setul{red}{2pt}
\setstcolor{red}

%



\maketitle

\begin{abstract}
Accurate and fast beam-alignment is essential to cope with the fast-varying environment in millimeter-wave communications.
A data-driven approach is a promising solution to reduce the training overhead by leveraging side information and on-the-field measurements.
In this work, a two-stage tensor completion algorithm is proposed to predict the received power on a set of possible users' positions, given received power measurements on a small subset of positions.
Based on these predictions {and on positional side information}, a small subset of beams is recommended to reduce the training overhead of beam-alignment.
Numerical results {evaluated with the Quadriga channel simulator} demonstrate that the proposed algorithm achieves correct alignment with high probability using small training overhead: given power measurement on only $20\%$ of the possible positions when using a discrete coverage area, our algorithm attains a probability of correct alignment of $80\%$, with only $2\%$ of trained beams, as opposed to a state-of-the-art scheme which achieves $50\%$ correct alignment in the same configuration.
To the best of our knowledge, this is the first work to consider the beam recommendation problem based on measurements collected on a small subset of positions.
\end{abstract}


%
\IEEEpeerreviewmaketitle

\vspace{-4mm}
\section{Introduction}

Millimeter wave (mmWave) and massive MIMO are the key technologies to enable high throughput communication in future wireless systems, with applications such as video-streaming, automated driving, cloud computing, etc \cite{6515173,7400949,6736761,6736746}.
However, narrow beams are required to compensate the path loss and severe signal propagation at the mmWave frequencies.
Narrow beam communication is especially challenging in mobile environments, since the beam direction needs to be continuously {trained}. 
Typically, this is achieved by sweeping over a finite set of candidate beamforming vectors to find the strongest beam direction \cite{Va2018,6600706}.
This process incurs huge overhead \cite{Akdeniz2014}
due to the potentially large set of candidate beamforming solutions that should be searched for in large antenna systems, calling for efficient beam-alignment protocols \cite{8573158}.

Beam-alignment has been a subject of intense research in recent years, with techniques ranging from beam-sweeping \cite{8573158}, angle of arrival and of departure (AoA/AoD) estimation \cite{6847111}, to data-assisted schemes \cite{Va2018}.
In particular, beam-sweeping schemes require to collect a set of beam measurements over the entire beam-space.
The simplest form of beam-sweeping is \emph{exhaustive search}, which scans through all possible beams between transmitter and receiver.
AoA/AoD estimation reduces the number of measurements by leveraging the sparsity of mmWave channels via compressive sensing \cite{6847111}.
In data-assisted schemes, mmWave channels are related to the environment of the user, such as its position, the geometry of the surrounding environment (e.g., buildings, vegetation, etc.) or temporal information (e.g., traffic).
Due to the difficulty to comprehensively model and accurately represent all propagation features in the environment, and how these affect mmWave propagation, a data-driven approach based on machine learning may be envisioned for this task.
In \cite{Va2018}, the authors proposed an inverse multi-path fingerprinting approach for beam-alignment utilizing prior measurements at a \emph{given position} to provide a set of candidate beam directions at the \emph{same} position.

{However, by requiring measurements to already be available at a certain position for predictions to be made, this approach fails to predict the channel in those positions where measurements are not yet available. For this reason, this scheme requires} to collect a huge amount of channel propagation measurements to cover the entire operational region, which may not be practical.
In many practical settings, the spatial correlation in the channel may be exploited to provide beam directions recommendations also in \emph{new} positions, where prior measurements are unavailable.
To address this more general problem, in this work we leverage the
tensor completion technique.
This problem has been recently investigated in many areas, such as computer vision, image in-painting, recommendation systems, etc. \cite{Cai_asingular}.

By exploiting the low-rank of mmWave MIMO channels \cite{7400949,6515173,6847111}, we construct a data model on a subset of positions as a tensor and formulate the tensor completion problem to estimate the channel on those positions {and beam directions} where measurements are missing.
To capture the channel spatial correlation, we introduce a smooth constraint that induces similarity among adjacent positions and beams.
We propose a two-stage tensor completion algorithm composed of two smooth matrix completions \cite{Cai_asingular} and a greedy selection algorithm to recommend a subset of candidate beams.
We show numerically that our proposed beam recommendation algorithm can provide accurate beam candidates with small training overhead.
Numerical evaluations demonstrate that our proposed method achieves $80\%$ probability of correct alignment with only $2\%$ of trained beams, given power measurements on only $20\%$ of discrete  random users' positions, as opposed to the state-of-the-art inverse-fingerprinting algorithm \cite{Va2018}, which attains a probability of correct alignment of $50\%$.

The rest of this paper is organized as follows.
In Sec. \uppercase\expandafter{\romannumeral2}, we present the system model and architecture;
in Sec. \uppercase\expandafter{\romannumeral3}, we propose the recommendation algorithm with two-stage tensor completion.
The numerical result are presented in Sec. \uppercase\expandafter{\romannumeral4},
followed by concluding remarks in Sec. \uppercase\expandafter{\romannumeral5}.

\vspace{-3mm}

\section{System Model}
In this section, we describe the channel model and the beamforming codebook of our communication system.
Then, we introduce the position-aided beam-alignment protocol and explain the goal of this work.
Later, we depict the data collection and the data tensor.
\vspace{-3mm}
\subsection{Channel Model}
\begin{figure}[t]
	\centering
	\includegraphics[width=8cm]{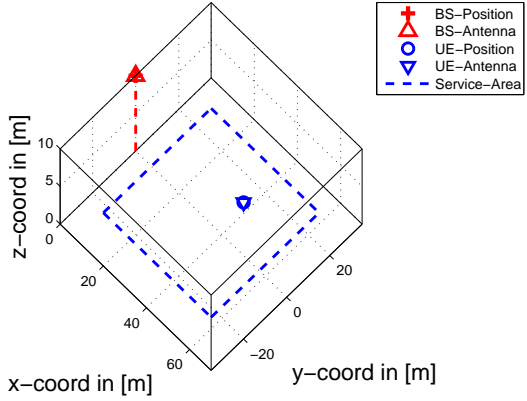}
	\caption{The network layout of the scenario.}
	\label{fig:Scenario}
\end{figure}

We consider a scenario with a base station (BS), servicing an area with GPS coordinates $\{(g_x,g_y): X_0 \leq g_x\leq X_{end},Y_0 \leq g_y\leq Y_{end}\}$, as in Fig. \ref{fig:Scenario}.
The geometric channel model \cite{7400949} is assumed for the uplink SIMO channel between the BS and the user (UE) at GPS coordinate $\mathbf{g}=(g_x,g_y)$ and is given by 
$$\mathbf{h}^{\mathbf{g}}=\sqrt{N_r}\sum^{L}_{\ell=1}{\alpha_\ell^{\mathbf{g}}}\mathbf{a}_r(\theta_{\ell}^{\mathbf{g}},\phi_{\ell}^{\mathbf{g}}),$$
where $\mathbf{a}_r(\theta_{\ell}^{\mathbf{g}},\phi_{\ell}^{\mathbf{g}})$  (see \eqref{atp}) is the normalized receive steering vector of the $\ell$th path; $\theta_{\ell}^{\mathbf{g}}$ and $\phi_{\ell}^{\mathbf{g}}$ are its elevation and azimuth angles;
$\alpha_\ell^{\mathbf{g}}$ is the complex channel gain; $L$ is the number of paths; and $N_r$ is the number of receive antennas.

In order to develop our data-driven approach, we aim to leverage the channel correlation with respect to some features of the environment where the UE is operating.
Here, we consider the correlation between the channel and the UE position.
To the best of our knowledge, the modeling of the propagation channel can only be achieved by real channel measurements or by simulation in ray-tracing software, which requires accurate modeling of the propagation environment, such as position of buildings, scatterers, etc.
In practice, our algorithm is applicable to a fully data-driven approach based on actual channel measurements. For evaluation purposes, in the numerical results in Sec. \ref{numerical_result}, we will generate these measurements with Quadriga
\cite{6758357}. 

\subsection{Beam Codebook and Received Signal Model}
We consider a uniform planar array (UPA) \cite{7400949,Va2018} at the BS with $N_x$ and $N_y$ antennas and $\lambda/2$ antenna spacing along the $x$ and $y$ directions (a total of $N_r=N_xN_y$ antennas), and a receive beamforming codebook
 $$\mathcal{W}{=}\{\mathbf w_{i,j}=\mathbf{a}(\theta_i,\phi_j),i=1,\cdots,C_{\theta},j=1,\cdots,C_{\phi}\},$$
of size $\vert\mathcal{W}\vert=C_\theta C_\phi$; $\mathbf{a}(\theta,\phi)$ is the array response vector representing a beam pointing in the elevation angle $\theta\in[-\pi/2,\pi/2)$ and the azimuth angle $\phi\in[-\pi/2,\pi/2)$,
\begin{equation}
\label{atp}
\begin{aligned}
    \mathbf{a}(\theta,\phi)=\frac{1}{\sqrt{N_r}}
    \begin{bmatrix}
           1\ e^{j\Omega_y}\ \cdots     e^{j({N}_y-1)\Omega_y}
    \end{bmatrix}^T
    \otimes\\
    \begin{bmatrix}
           1\ e^{j\Omega_x}\ \cdots e^{j({N}_x-1)\Omega_x}
    \end{bmatrix}^T,
\end{aligned}
\end{equation}
with $\Omega_y{=}{\pi}\sin{\theta}\sin{\phi}$, $\Omega_x{=}{\pi}\sin{\theta}\cos{\phi}$.
To construct $\mathcal W$, $\theta_i$ and $\phi_j$ are uniformly quantized in $[-\pi/2,\pi/2)$ with resolution $\pi/C_{\theta}$ and $\pi/C_\phi$ as
\begin{align}
 &   \theta_i = -\frac{\pi}{2}+(i-1)\times\frac{\pi}{C_{\theta}}, \ i =1,\cdots,C_{\theta},\\
  &  \phi_j = -\frac{\pi}{2}+(j-1)\times\frac{\pi}{C_{\phi}}, \ j =1,\cdots,C_{\phi}.
\end{align}
We index the beamforming vectors in $\mathcal W$ as
$$\mathcal I\equiv\{(i,j):i =1,\cdots,C_{\theta},j =1,\cdots,C_{\phi}\}.$$

We consider an uplink beam training scheme, in which the UE at GPS coordinate $\mathbf{g}=(g_x,g_y)$ transmits a known unit-norm training sequence vector $\mathbf{s}\in\mathbb{C}^{N\times 1}$
and the BS processes the received signal with the beamforming vector $\mathbf w_{i,j}$, yielding the received signal vector
$$\mathbf y_{i,j}^{\mathbf{g}}=\sqrt{P_t}\mathbf{w}_{i,j}^H\mathbf{h}^{\mathbf{g}} \mathbf s + \mathbf{v},$$
where $\mathbf{v}\sim\mathcal{CN}(0,\sigma_v^2 \mathbf I)$ is the received noise vector; 
$P_t$ is the transmit power.
The received power with the $({i,j})$-th beamformer can be estimated as
$$r_{i,j}^{\mathbf{g}} =
 |\mathbf s^{H}\mathbf y_{i,j}^{\mathbf{g}}|^2=|\sqrt{P_t}\mathbf{w}_{i,j}^H\mathbf{h}^{\mathbf{g}}+\Tilde{v}|^2,$$
where $\Tilde{v}=\mathbf s^{H}\mathbf v$ is zero-mean complex Gaussian noise with variance $\sigma_v^2$.
These received powers are stored in a database explained in Sec. \ref{data_model_sec}, and then used in our completion framework to predict the received power in other positions and beams.

\begin{figure}[t]
	\centering
	\includegraphics[width=7cm]{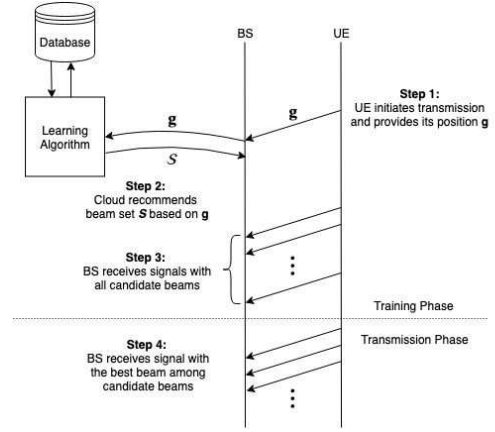}
	\caption{Position-aided beam alignment protocol.}
	\label{fig:BF_recommendation}
\end{figure}

\vspace{-4mm}
\subsection{Position-Aided Beam-Alignment}
The idea of this approach is to provide a set of candidate beams at a given UE position.
Since the overhead of the conventional beam-sweeping approach is unacceptable (it scales with $|\mathcal W|$ and is typically very large), our objective is to design a learning algorithm that provides a small subset $\mathcal S\subset\mathcal W$ of candidate beams for training, which is likely to contain the best beam (the one with highest received power).

In Fig. \ref{fig:BF_recommendation}, we introduce a flow diagram for the position-aided beam-alignment.
In step 1, the UE initiates the uplink mmWave transmission request accompanied with its GPS coordinate $\mathbf{g}=(g_x, g_y)$ to the BS using sub-6GHz control channels.
The position information is available via a suite of sensors such as GPS or LIDAR \cite{7535489,Va2018}.
In step 2, the BS forwards the UE's GPS coordinate $\mathbf{g}$ to the cloud, which then processes the learning algorithm and provides the recommended beam set $\mathcal S$.
In step 3, the UE transmits a sequence of $|\mathcal{S}|$ known signals, and the BS receives these signals with the codewords in the recommended beam set $\mathcal S$.
Then, the BS selects the best beam (ranked by received power) among the recommended beams.
In step 4, the BS uses the selected receive beamforming vector for the subsequent mmWave uplink data transmission.
Due to the reciprocity of the wireless channel, the BS can also utilize these beam directions for downlink transmissions.

\vspace{-4mm}

\subsection{Data Model}
\label{data_model_sec}
Data is an essential element for the machine learning approach. 
Here, we describe how we store information in the database.
We discretize the service area of the BS with resolution $\Delta_s$ and define the position labels $\mathbf{p}=(p_x,p_y)$ as the function of GPS coordinate $\mathbf g$,
\begin{equation}
\label{pos_label_derive}
\mathbf{p}(\mathbf{g})=\left(1 + \round{\frac{g_x-X_0}{\Delta_s}},1 + \round{\frac{g_y-Y_0}{\Delta_s}}\right),
\end{equation}
where $\round{x}$ denotes the nearest integer to $x$; the x-axis label $p_x\in \{1,\cdots,L_x\}$ where $L_x = \left\lceil\frac{X_{end}-X_0}{\Delta_s}\right\rceil$; and the y-axis label $p_y\in \{1,\cdots,L_y\}$ where $L_y =\left\lceil \frac{Y_{end}-Y_0}{\Delta_s}\right\rceil$.
The BS measures the received power based on a fixed transmit power $P_t$.
We define the received power with beam $\mathbf{w}_{i,j}$ at position $\mathbf{p}$ as ${r}^{(\mathbf{p}(\mathbf g),i,j)}=r_{i,j}^{\mathbf{g}}$.
During the data collection, the BS might collect multiple measurements for the same beamforming {vector} and position.
For beam $\mathbf{w}_{i,j}$ and position $\mathbf p$, we define ${r}^{(\mathbf p,i,j)}_k$ as the $k$-th measured received power and $N_{ob}^{(\mathbf p,i,j)}$ as the number of measurements collected so far on that position and beam index.
We extract the average power information by computing the sample mean $\Bar{r}^{(\mathbf p,i,j)} = \frac{1}{N_{ob}^{(\mathbf p,i,j)}}\sum_{k=1}^{N_{ob}^{(\mathbf p,i,j)}}r^{(\mathbf{p},i,j)}_k.$
Once the BS performs a new measurement for position $\mathbf p$ and beam $\mathbf w_{i,j}$, we can update the average received power {in an online fashion} as
$$\Bar{r}^{(\mathbf p,i,j)}\leftarrow \frac{N-1}{N}\Bar{r}^{(\mathbf p,i,j)} + \frac{1}{N}r^{(\mathbf{p},i,j)}_{N},$$
where $N=N_{ob}^{(\mathbf p,i,j)}+1$, followed by $N_{ob}^{(\mathbf p,i,j)}\leftarrow N_{ob}^{(\mathbf p,i,j)}+1$.
Then, the database records the average received power along with the side information, including the UE's position $\mathbf{p}=(p_x,p_y)$, and the indices of the beamforming codeword $\mathbf{w}_{i,j}$, as in TABLE \ref{table:database}.

\begin{table}[ht]
	\caption{Database form} 
	\centering 
	\begin{tabular}{c c c c | c } 
		\hline
		\hline 
		$p_x$ & $p_y$ & $i$ & $j$  & $\bar{r}^{(\mathbf p,i,j)}$ \\ [0.5ex] 
		\hline 
		1 & 1  & 1 & 4  & 5.2 \\
		1 & 2  & 4 & 5  & 6.1\\
		$\vdots$ & $\vdots$ & $\vdots$ & $\vdots$  & $\vdots$\\ [1ex] 
		\hline 
		\hline 
	\end{tabular}
	\label{table:database} 
		\vspace{-1.8mm}
\end{table}

We represent the extracted data as a $4$-th order tensor
\begin{equation}
\label{database}
\begin{aligned}
        &\mathcal{T}(p_x,p_y,i,j)=
        \left\{
            \begin{array}{ll}
            \Bar{r}^{(p_x,p_y,i,j)}, & (p_x,p_y,i,j)\in\Psi,\\
             0, & \text{otherwise,}\\
             \end{array}
        \right.
\end{aligned}  
\end{equation}
where $\Psi$ is the set of observed combinations of positions and beams stored in the database and the unobserved entries $(p_x,p_y,i,j)\notin\Psi$ are set to zero.
It is impractical to collect the information with all combinations of positions and beam-directions into the database due to the limited sampling resources. 
For this reason, some positions possibly have no representation in the database.
Even in the observed positions, there might be only a limited number of beams' information recorded.
Therefore, the tensor $\mathcal{T}$ may be highly incomplete.
\vspace{-3mm}


\section{Tensor completion and Beam recommendation}
Our goal is to recommend a set $\mathcal S$ of $N_{tr}$ candidate beams for the UE to train based on its position.
If the UE is in a position $\mathbf p$ represented in the database, and with all beam measurements available, $\Bar r^{(p_x,p_y,i,j)},\forall (i,j)$, then this task can be easily accomplished by recommending the $N_{tr}$ beams  with highest average received power in the given position.

Otherwise, we design an algorithm based on tensor completion that employs the knowledge at neighboring positions to support the beam recommendation for the UE.
Note that successful completion is highly dependent on the sampling set.
If measurements are missing on a certain row or column of an incomplete matrix, then no reconstruction is possible on that row or column, if we only rely on a low-rank approximation \cite{Cai_asingular}.
A similar issue also exists in the tensor case:
if no measurements are available on a certain index in a given dimension, no elements corresponding to this index can be predicted by only relying on low rank structure.
In our problem, the database contains measurements related to a subset of beams on few observed positions.
The data tensor might be so incomplete that we cannot guarantee that every index of each dimension is measured at least once.
Therefore, the low-rank tensor completion often fails to provide predictions on unobserved positions.

To address this challenge, in addition to the low-rank approximation, we enforce a smoothness constraints across adjacent entries on a given dimension. This constraint captures realistic spatial correlations across adjacent positions arising in mmWave channels: 
similarity between neighboring beams at a given position;
and similarity between neighboring positions on a given beam direction. 
{In Sec. \ref{2stageTC}, we propose a two-stage tensor completion implemented by dividing the tensor completion into two smooth matrix completions (SMCs), considered in Sec. \ref{SMC}.}
Finally, we propose a greedy algorithm to provide the recommended beams based on the predicted received power in Sec. \ref{recalg}.

\vspace{-4.5mm}

\subsection{Two-stage Tensor Completion}
\label{2stageTC}
Given the data tensor $\mathcal{T}$ in \eqref{database} and the  set  of  observed  combinations  of  positions and  beams $\Psi$, we aim to recover the incomplete tensor ${{\mathcal{T}}}$.
Since the tensor $\mathcal{T}$ is highly incomplete, we might only have limited number of beams' information on few observed positions, causing the low-rank completion to fail. To address this challenge, we propose a two-stage tensor completion, each based on SMC.

In the first stage, for each observed position $(x_o,y_o)$ such that $(x_o,y_o,i,j)\in\Psi$ for some $(i,j)$, we do the SMC on the beam matrix to predict the received power on the unobserved beams, by exploiting the low-rank property that the received powers of beams in a given position tend to concentrate in few beam clusters due to the limited scattering of mmWave channels.
The smoothness between neighboring beams depends on the beamwidth of the receive beamforming and the angular spread of the channel \cite{doi:10.4218/etrij.2017-0188}.
To this end, let $\mathbf{B}^{(x_o,y_o)}=\mathcal{T}(x_o,y_o,:,:)$ be the (possibly incomplete) matrix of received powers along beam directions, for the given position $(x_o,y_o)$; 
let $\Omega\equiv\{(i,j):(x_o,y_o,i,j)\in\Psi\}$ be the set of observed beams in position $(x_o,y_o)$. 
Then, the SMC can be expressed as
\begin{align}
\nonumber
    &\textbf{SMC}_{\Omega}(\mathbf{B}^{(x_o,y_o)} ){=}\arg\min_{\mathbf X}\lVert \mathbf{X} \rVert_*\!{+} \gamma(\lVert \mathbf{D}_{C_{\theta}} \mathbf{X} \rVert_F^2{+}\lVert \mathbf{X} \mathbf{D}_{C_\phi}^{T}\rVert_F^2)\\
    &\text{s.t. }\mathbf{X}_\Omega=\mathbf{B}^{(x_o,y_o)}_\Omega
    \label{SMC_B}
\end{align}
and is considered in Sec. \ref{SMC}.
The first term of the objective function is the nuclear norm 
of $\mathbf X$, $\lVert \mathbf{X} \rVert_* = \sum_{i=1}^{\min(m,n)}\sigma_i{,}$
where $\sigma_i$ is the $i$-th largest singular value of $\mathbf{X}$.
The second term of the objective function is a penalty term which induces smoothness across entries in each row and column of $\mathbf X$.
The matrix $\mathbf{D}_m\in \mathbb{R}^{(m-1)\times m}$ is the smoothness matrix, capturing the differences between neighboring entries of a matrix:
\begin{equation}
    \mathbf{D}_m=
    \begin{bmatrix}
    1 & -1 & \cdots & 0 & 0 \\
    \vdots & 1 & -1 & \vdots & 0 \\
    0 & \vdots & \ddots & \ddots & \vdots \\
    0 & 0 & \cdots & 1 & -1 
    \end{bmatrix}_{(m-1)\times m}.
\end{equation}
Thus, $\lVert \mathbf{X} \mathbf{D}^T_n \rVert_F^2$ and $\lVert \mathbf{D}_m \mathbf{X} \rVert_F^2$ quantify the row and column smoothness of matrix $\mathbf{X}$, respectively.
In our optimization, we consider smoothness on rows and columns simultaneously,
as opposed to LTVNN \cite{XuHan2014}, which considers them separately.

After the first stage completion by {SMC}, we obtain the data tensor ${\mathcal{T}'}$ completed on the observed positions and update $\Psi'$ by setting all predicted terms as observed.
In the second stage, the SMC is implemented on the position matrix for each beam, by leveraging the fact that the received power tends to vary smoothly between neighboring positions on a given beam.
For a given beam, the smoothness between neighboring positions is related to the position resolution $\Delta_s$.
Specifically, in each beam $(i_o,j_o)$ such that $(p_x,p_y,i_o,j_o){\in}\Psi'$ for some $(p_x,p_y)$,
let $\mathbf{G}^{(i_o,j_o)}{=}{\mathcal{T}'}(:,:,i_o,j_o)$ be the (possibly incomplete) matrix of received powers on different positions, along the given beam indexed by $(i_o,j_o)$; 
let $\Omega\equiv\{(p_x,p_y):(p_x,p_y,i_o,j_o)\in\Psi'\}$ be the set of observed positions along the beam $(i_o,j_o)$. 
Then, the SMC can be expressed as
\begin{align}
\nonumber
    &\textbf{SMC}_{\Omega}(\mathbf{G}^{(i_o,j_o)}){=}\arg\min_{\mathbf X}\lVert \mathbf{X} \rVert_*{+}\gamma(\lVert \mathbf{D}_{L_x} \mathbf{X} \rVert_F^2{+}\lVert \mathbf{X} \mathbf{D}_{L_y}^{T}\rVert_F^2)\\
    &\text{s.t. }\mathbf{X}_\Omega=\mathbf{G}^{(i_o,j_o)}_\Omega
        \label{SMC_M}
\end{align}
and is considered in Sec. \ref{SMC}.
After the second stage completion by SMC, we get the completed tensor $\hat{\mathcal{T}}$, which predicts the received power of all unknown positions/beams, and is used in Sec. \ref{recalg} to recommend the set of training beams at a given UE position $\mathbf p$.
The two-stage tensor completion algorithm is shown in \textbf{Algorithm \ref{Two_stage_TC}}.
Next, we discuss the solutions of the SMC problems \eqref{SMC_B} and \eqref{SMC_M}, solved by \textbf{Algorithm \ref{SMC_alg}}.

\vspace{-4mm}
\subsection{Smooth Matrix Completion (SMC)}
\label{SMC}
The smooth matrix completion exploits both the low rank and the smoothness of the data.
Given the incomplete matrix $\mathbf{M}{\in}\mathbb{R}^{m\times n}$ with $\mathbf{M}_{ij},\forall (i,j){\in}\Omega$, the SMC problem to predict the unobserved entries $\mathbf{M}_{ij},\ \forall (i,j){\notin}{\Omega}$ is expressed as
\begin{equation}
    \min_{\mathbf X} \ \lVert \mathbf{X} \rVert_* + \gamma(\lVert \mathbf{D}_m \mathbf{X} \rVert_F^2+\lVert \mathbf{X} \mathbf{D}_n^{T}\rVert_F^2)\ 
    \text{s.t. }\mathbf{X}_\Omega=\mathbf{M}_\Omega,
\label{opt_prob_1}
\end{equation}
where $\gamma$ is a regularization parameter.
We use the alternating direction method of multipliers (ADMM) \cite{Yang_linearizedaugmented} to efficiently solve (\ref{opt_prob_1}).
With ADMM, we reformulate the problem as
\begin{equation}
\label{opt_prob_2}
\begin{aligned}
    \min_{\mathbf{X},\mathbf{Y}} \ & \lVert \mathbf{X} \rVert_*{+}\gamma(\lVert \mathbf{D}_m \mathbf{Y} \rVert_F^2{+}\lVert \mathbf{Y} \mathbf{D}_n^{T}\rVert_F^2)+\frac{\lambda}{2}\lVert \mathbf{Y}-\mathbf{X} \rVert_F^2\\
    \text{s.t. }&\mathbf{Y}_\Omega=\mathbf{M}_\Omega{,}\ \ 
                \mathbf{X}=\mathbf{Y}{,}
\end{aligned}
\end{equation}
where $\lambda>0$ is a small fixed parameter.
We introduce the Lagrangian multiplier $\mathbf{Z}$ associated with the constraint ${\mathbf{X}=\mathbf{Y}}$.
The augmented Lagrange function of \eqref{opt_prob_2} is 
\begin{equation}
\begin{aligned}
    L(\mathbf{X,Y,Z})=& \lVert \mathbf{X} \rVert_* + \gamma (\lVert \mathbf{D}_m \mathbf{Y} \rVert_F^2+\lVert \mathbf{Y} \mathbf{D}_n^{T}\rVert_F^2)\\
    &+{\mathrm{tr}}(\mathbf{Z}^T(\mathbf{Y}-\mathbf{X}))+\frac{\lambda}{2}\lVert \mathbf{Y}-\mathbf{X} \rVert_F^2.
\end{aligned}
\end{equation}
The ADMM algorithm is implemented by minimizing iteratively $L(\mathbf{X,Y,Z})$ over $\mathbf{X}$ and $\mathbf{Y}$, and then update $\mathbf{Z}$ as
\begin{equation}
\label{update_overview}
\begin{aligned}
&\mathbf{X}_{t+1}=\arg\min_{\mathbf{X}}L(\mathbf{X},\mathbf{Y}_t,\mathbf{Z}_t);\\
&\mathbf{Y}_{t+1}=\arg\min_{\mathbf{Y}}L(\mathbf{X}_{t+1},\mathbf{Y},\mathbf{Z}_t)\text{, s.t. }\mathbf{Y}_\Omega=\mathbf{M}_\Omega;\\
&\mathbf{Z}_{t+1}=\mathbf{Z}_{t}+{\beta}(\mathbf{Y}_{t+1}-\mathbf{X}_{t+1});
\end{aligned}
\end{equation}
where $\beta$ is a step-size.
To optimize $\mathbf X$, we minimize $L(\mathbf{X,Y,Z})$ with fixed $\mathbf{Y}_t$ and $\mathbf{Z}_t$, yielding 
$$\mathbf X_{t+1}=\arg\min_{\mathbf{X}} \lVert \mathbf{X} \rVert_* + \frac{\lambda}{2}\Big\lVert \mathbf{X}-(\mathbf{Y}_t+\frac{1}{\lambda}\mathbf{Z}_t)\Big\rVert_F^2.$$
In \cite{Cai_asingular}, it is shown that this problem is strictly convex and its solution is given by the singular value thresholding, $\mathbf{X}_{t+1}= \mathcal{D}_{1/\lambda}\left(\mathbf{Y}_{t}+\frac{1}{\lambda}\mathbf{Z}_{t}\right){,}$
where $\mathcal{D}_\tau$ is the soft-thresholding operator.
For a matrix $\mathbf A$ with singular value decomposition (SVD) $\mathbf{A}=\mathbf{U}\mathbf{\Sigma}\mathbf{V}^H$,
where $\mathbf{\Sigma}=\text{diag}(\sigma_1,\dots\sigma_r)$, this is defined as
$\mathcal{D}_\tau(\mathbf{A})=\mathbf{U}\mathcal{D}_\tau(\mathbf{\Sigma})\mathbf{V}^H,\ \mathcal{D}_\tau(\mathbf{\Sigma})=\text{diag}(\{\max\{\sigma_i-\tau,0\}\}).$

The minimization of $L(\mathbf{X,Y,Z})$ over $\mathbf Y$ with fixed $\mathbf{X}_{t+1}$ and $\mathbf{Z}_t$ can be formulated as
\begin{equation}
\label{optprobY}
\begin{aligned}
    \mathbf Y_{t+1}=&\arg\min_{\mathbf{Y}}\ \gamma (\lVert \mathbf{D}_m \mathbf{Y} \rVert_F^2+\lVert \mathbf{Y} \mathbf{D}_n^{T}\rVert_F^2)\\
    &\hspace{5mm}+{\mathrm{tr}}(\mathbf{Z}_{t}^T(\mathbf{Y}-\mathbf{X}_{t+1}))+\frac{\lambda}{2}\lVert \mathbf{Y}-\mathbf{X}_{t+1} \rVert_F^2\\
    &\text{s.t. }\mathbf{Y}_\Omega=\mathbf{M}_\Omega.
\end{aligned}
\end{equation}
To solve this problem, we restrict its optimization to the unobserved set
$\bar{\Omega}\equiv\{(i,j):(i,j)\notin\Omega\}$, and force $\mathbf Y_{ij}=\mathbf M_{ij}$ for $(i,j)\in\Omega$.
By computing the derivative of \eqref{optprobY} with respect to $\mathbf Y_{ij}$ for $(i,j)\in\bar\Omega$ and setting it to zero, we obtain the equation $\text{tr}\left(\{\mathbf{U}^{(i,j)}\}^T\mathbf{Y}\right)-\frac{\lambda \mathbf X_{ij} - \mathbf Z_{ij}}{2\gamma}=0$,
where
\begin{align*}
\mathbf{U}^{(i,j)}=&\mathbf{D}_m^T\mathbf{D}_m \mathbf{e}_{m}(i) \mathbf{e}_{n}(j)^T + \mathbf{e}_{m}(i) \mathbf{e}_{n}(j)^T \mathbf{D}_n^T\mathbf{D}_n\\& + \frac{\lambda}{2\gamma} \mathbf{e}_{m}(i) \mathbf{e}_{n}(j)^T.
\end{align*}
The vector $\mathbf{e}_{m}(i)$ is an $m\times 1$ vector with the $i$-th element equal to $1$ and all other elements equal to $0$.
We force $\mathbf{Y}_\Omega = \mathbf{M}_\Omega$ as in \eqref{optprobY}, yielding the set of 
$|\bar\Omega|$ linear equations, $\forall (i,j)\in\bar\Omega$,
\begin{align}
\label{Y_equation}
&\sum_{(p,q)\in\bar\Omega}\mathbf U^{(i,j)}_{pq}\mathbf Y_{pq}
=\frac{\lambda \mathbf X_{ij} - \mathbf Z_{ij}}{2\gamma}-\sum_{(p,q)\in\Omega}\mathbf U^{(i,j)}_{pq}\mathbf M_{pq}.
\end{align}
There are $\vert \bar\Omega \vert$ unknowns $\left(\mathbf Y_{pq},\ (p,q)\in{\bar\Omega}\right)$ and $\vert {\bar\Omega} \vert$ linear equations, so the matrix $\mathbf{Y}_{\bar\Omega}$ can be derived by solving \eqref{Y_equation}.
Then, we update the Lagrangian multiplier $\mathbf{Z}$ with fixed $\mathbf{X}_{t+1}$ and $\mathbf{Y}_{t+1}$ as in \eqref{update_overview}.
The algorithm updates $\mathbf{X}$, $\mathbf{Y}$, and $\mathbf{Z}$ iteratively until 
a stop criterion is satisfied.
With the convergence of ADMM \cite{Boyd:2011:DOS:2185815.2185816}, the iteration approaches the feasibility $\Vert\mathbf{X}_t-\mathbf{Y}_t\Vert_F\rightarrow 0$, thus we set the stop condition as $\Vert\mathbf X_t-\mathbf Y_t\Vert_F\leq\epsilon$.
It follows that $\Vert\mathbf{Z}_{t+1}-\mathbf{Z}_t\Vert_F=\beta\Vert\mathbf Y_{t+1}-\mathbf X_{t+1}\Vert_F\rightarrow 0$, which guarantees the convergence of $\mathbf Z$.
{The computational complexity of SMC is dominated by the SVD {to perform the} soft-thresholding {operation in each} iteration, which is $\mathcal O(mn^2)$ for $m\geq n$ \cite{GoluVanl96}.
The update of $\mathbf{Y}$ requires a matrix inversion per iteration, whose computational complexity is $\mathcal O(\lvert\bar\Omega\rvert^3)$.}

\begin{algorithm}[]
	\caption{Two-stage Tensor Completion}
	\begin{algorithmic}[1]
		\INPUT incomplete data tensor $\mathcal{T}$ and the observed set $\Psi$
		\OUTPUT $\hat{\mathcal{T}}$
		\FOR{$(x_o,y_o,-,-)\in\Psi$}
		\STATE $\mathbf{B}^{(x_o,y_o)}
		=\mathcal{T}(x_o,y_o,:,:)$
		\STATE $\Omega = \{(i,j)\vert(x_o,y_o,i,j)\in\Psi\}$
		\STATE $\mathcal{T}'(x_o,y_o,:,:)=\textbf{ SMC}_{\Omega}(\mathbf{B}^{(x_o,y_o)})$
		\STATE Update the observed set ${\Psi'}$
		\ENDFOR
		\FOR{$(-,-,i_o,j_o)\in\Psi'$}
		\STATE $ \mathbf{G}^{(i_o,j_o)}={\mathcal{T}'}(:,:,i_o,j_o)$
		\STATE $\Omega = \{(p_x,p_y)\vert(p_x,p_y,i_o,j_o)\in\Psi'\}$
		\STATE $\hat{\mathcal{T}}(:,:,i_o,j_o)=\textbf{ SMC}_{\Omega}(\mathbf{G}^{(i_o,j_o)})$
		\ENDFOR
		
	\end{algorithmic}
	\label{Two_stage_TC}
\end{algorithm}
	\vspace{-1mm}
\begin{algorithm}[]
	\caption{Smooth Matrix Completion (SMC)}
	
	\begin{algorithmic}[1]
		\INPUT incomplete data matrix $\mathbf{M}$ and  observed set $\Omega$
		\OUTPUT $\hat{\mathbf{M}}$
		\STATE \textbf{Initialization} 
		{$\mathbf X_t=\mathbf Y_t=\mathbf M$,
			$\mathbf Z_t=\mathbf 0$, $\epsilon_t=\infty$}
		\WHILE{$\epsilon_t>\epsilon$}
		\STATE $\mathbf{X}_{t+1}= \mathcal{D}_{1/\lambda}\left(\mathbf{Y}_{t}+\frac{1}{\lambda}\mathbf{Z}_{t}\right)$
		\STATE $\mathbf{Y}_{t+1} =\mathbf{Y}_{\bar\Omega}+\mathbf{M}_\Omega$ ($\mathbf{Y}_{\bar\Omega}$ {obtained}  by solving \eqref{Y_equation})
		\STATE $\mathbf{Z}_{t+1}=\mathbf{Z}_{t}+\beta(\mathbf{Y}_{t+1}-\mathbf{X}_{t+1})$
		\STATE $\epsilon_{t+1}=\Vert\mathbf{X}_{t+1}-\mathbf{Y}_{t+1}\Vert_F$; $t:=t+1$
		\ENDWHILE
		\STATE $\hat{\mathbf{M}}\leftarrow \mathbf{X}_{{\bar\Omega}}+\mathbf{M}_{\Omega}$
	\end{algorithmic}
	\label{SMC_alg}
\end{algorithm}

\subsection{Recommendation Algorithm}
\label{recalg}
With the completed tensor ${\hat{\mathcal{T}}}$, we have the estimated received powers of all beams at UE position $(p_x,p_y)$.
Suppose the number of beams to be trained is $N_{tr}$, the construction of the recommended beam set is a beam subset selection problem.
It can be directly fulfilled by selecting the $N_{tr}$ beams with largest predicted received power from the completed tensor ${\hat{\mathcal{T}}}$ using \textbf{Algorithm \ref{Beam_subset_selection}}.


\begin{algorithm}
	\caption{Beam Subset Selection}
	\begin{algorithmic}[1]
		\INPUT completed tensor $\hat{\mathcal{T}}$, beam number $N_{tr}$, beam codebook $\mathcal{W}$ with indices $\mathcal I$, UE position $(p_x,p_y)$
		\OUTPUT recommended beam subset ${\mathcal S}_{N_{tr}}$
		\STATE \textbf{Initialization} ${\mathcal S}_0\leftarrow\emptyset$
		\FOR{$n=1:N_{tr}$}
		\STATE $(i^*,j^*)=\arg\max_{(i,j)\in \mathcal{I}\setminus\mathcal S_{n-1}}\hat{\mathcal{T}}({p}_x,{p}_y,i,j)$
		\STATE $\mathcal {S}_n\leftarrow \mathcal {S}_{n-1}\cup {(i^*,j^*)}$
		\ENDFOR
	\end{algorithmic}
	\label{Beam_subset_selection}
\end{algorithm}

\vspace{-4mm}

\section{Numerical Results}
\label{numerical_result}
Here, we evaluate the misalignment probability and the spectral efficiency of our proposed beam recommendation algorithm (\textbf{TC}) with the channel generated by Quadriga \cite{6758357}.

\vspace{-4.3mm}

\subsection{Experiment Setting}
\vspace{-0.5mm}
We consider an uplink SIMO scenario, with an UPA having $N_x=N_y=16$ antennas along the $x$ and $y$ directions (as in \eqref{atp}) at the BS, and isotropic antenna at the UE.
The scenario \textit{mmMAGIC\_UMi\_NLOS} is selected, with carrier frequency $f_c=58.68$ GHz.
The UPA codebook size is $\vert\mathcal{W}\vert=256$ with $(C_\theta,C_\phi)=(16,16)$.
The network layout is {depicted} in Fig. \ref{fig:Scenario}, containing one BS at $(0,0,10)$ serving the UE in the area $\mathcal A=\{(g_x,g_y):10 \leq g_x\leq 60 ,-25 \leq g_y\leq 25 \}$ with height as $1.5$ m.
Considering $51\times 51=1261$ reference GPS coordinates uniformly located in the service area $\mathcal A$, we collect the SIMO channel for each reference GPS coordinate in $\mathcal A$ as the ground truth data.
The position labels of $\mathcal A$ are derived as in \eqref{pos_label_derive} with the resolution $\Delta_s = 5$m, where the length are $L_x=11$ and $L_y=11$.
The data tensor can be expressed by $\mathcal{T}\in \mathbb{R}^{L_x \times L_y \times C_\theta \times C_\phi}$, where $(L_x,L_y,C_\theta,C_\phi)=(11,11,16,16)$.
The observed position ratio $K_{op}=C_{op}/(L_xL_y)$ is varied, where $C_{op}$ denotes the number of observed positions.
Regarding the observed set of data tensor, we make the two following assumptions for the experiment setting.
\textbf{Assusmption 1:} The observed positions are randomly chosen. For the observed position $\mathbf p'$, the measurements of the reference GPS coordinates corresponding to position $\mathbf p'$, $\{\mathbf g =(g_x,g_y): \mathbf{p}(\mathbf{g})=\mathbf{p}',\ \mathbf{g}\in \mathcal A\}$, are observed.
\textbf{Assusmption 2:} For each observed GPS coordinate $\mathbf{g}$, only the measurements of the top $10\%$ beams (ranked by received power) are stored in the database. 
With these two assumptions, the data tensor in \eqref{database} is incomplete in both positions' and beams' dimensions.

Our formulation allows to make predictions for unknown beams/positions by exploiting spatial correlation. On the other hand, the previous work \cite{Va2018} for mmWave beam alignment uses the prior knowledge already available at a given position, but does not allow to make predictions if measurements are not available.
For comparison, we consider the type B fingerprinting method \cite{Va2018} by providing the recommended beam set based on the closest position having available prior knowledge, if the prior measurements of UE position are not given.

\vspace{-3.2mm}
\subsection{Performance of Proposed Beam-alignment Algorithm}
\begin{figure}[t]
	\centering
	\includegraphics[width=6.6cm]{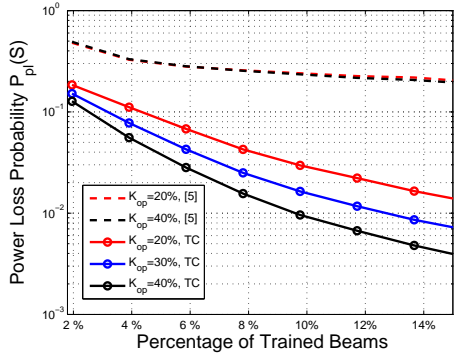}
	\caption{Power loss probability versus the percentage of trained beams.}
	\label{fig:Exp1}
\end{figure}

We evaluate the power loss probability  $P_{pl}({\mathcal S}_{{\mathbf p}})$  versus the percentage of trained beams $(\vert\mathcal{S}_{{\mathbf p}}\vert/\vert\mathcal{W}\vert)$ under different observed position ratio $K_{op}$.
The noise impact is ignored.
To measure the beam alignment accuracy for the recommended set ${\mathcal S_{\mathbf p}}$, we define the metric as the power loss probability $P_{pl}({\mathcal S_{\mathbf p}}) = 1-P_{s}({\mathcal S_{\mathbf p}})$, 
where $P_{s}({\mathcal S_{\mathbf p}})$ is the probability that the best beam is included in the set $\mathcal{S}_{\mathbf p}$ at position ${\mathbf p}$,
\begin{equation}
    P_{s}({\mathcal S_{\mathbf p}}) = \mathbb{P}\left(\max_{{(i,j)}\in\mathcal{S}_{\mathbf p}}r^{(\mathbf{p},{i,j})} = {\max_{{(i',j')\in\mathcal{I}}}r^{(\mathbf{p},{i',j'})}}\right).
\end{equation}
We average the power loss probability over the channels at the GPS coordinates corresponding to unobserved positions.
In Fig. \ref{fig:Exp1}, $P_{pl}(\mathcal S_{\mathbf p})$ decreases when the database contains more observed positions with fixed number of trained beams.
The type-B fingerprinting method \cite{Va2018} requires at least $13\%$ of the beams to be trained to attain $P_{pl}(\mathcal S_{\mathbf p})=20\%$.
With $K_{op}$ increasing from $20\%$ to $40\%$, $P_{pl}(\mathcal S_{\mathbf p})$ only very slightly improves.
For our proposed method (\textbf{TC}), to attain $P_{pl}(\mathcal S_{\mathbf p})=20\%$, we require only $2\%$ trained beams when $K_{op}=20\%$.
With more trained beams or higher $K_{op}$, $P_{pl}(\mathcal S_{\mathbf p})$ of \textbf{TC} decreases significantly.

\vspace{-3.5mm}

\subsection{Spectral Efficiency}
We evaluate the average spectral efficiency versus the average transmit power.
We first define the transmission rate as 
\begin{equation}
    R=B\log_2(1+\eta P_t \Vert \mathbf{w}^H\mathbf{h}\Vert^2)
\end{equation}
where $B=1.76$ GHz is the bandwidth\cite{Va2018,6758357}{;} $P_t$ is the transmit power; $\mathbf{w}$ is the selected beamforming vector; $\mathbf{h}$ is the SIMO channel; $\eta\triangleq\frac{\Lambda^2\zeta}{8\pi d^2 N_0 B}$ is the SNR scaling factor, where 
$\Lambda=c/f_c$ is the wavelength ($c$ is the speed of light), $N_0=-174$ dBm/Hz is the noise power spectral density, $d$ is the distance, and $\zeta=1$ is the antenna efficiency.
The microslot duration $\delta_S=10\ \mu s$ is the time for one transmission.
The frame time $T_{frame}=5 $ ms is fixed.
The training time is $T_{train}=N_{tr}\times\delta_S$.
The fraction of time used for data transmission is 
$f_{comm} = \frac{T_{frame}-T_{train}}{T_{frame}}$.
Then, the average throughput is $\Bar{R}=R\times f_{comm}$.

\begin{figure}[t]
    \centering
    \includegraphics[width=6.6cm]{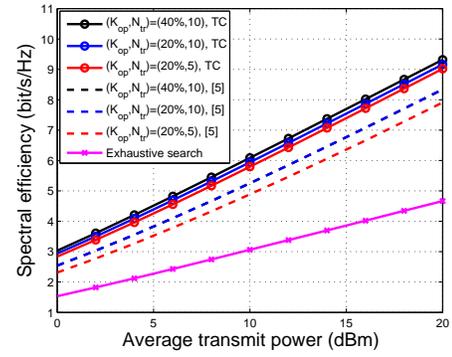}
    \caption{Spectral efficiency $(\Bar{R}/B)$ versus the average transmit power $({P_t})$. 
    }
    \label{fig:Exp2}
\end{figure}

In Fig. \ref{fig:Exp2}, the trend of spectral efficiency $(\Bar{R}/B)$ is monotone increasing.
We compare our proposed scheme (\textbf{TC}) with the type-B fingerprinting method \cite{Va2018} and the \emph{exhaustive search} which trains all ${|\mathcal W|}$ beams in the training phase.
\textbf{TC} is better than \emph{exhaustive search}.
The spectral efficiency of \textbf{TC} with $(K_{op},N_{tr})=(40\%,10)$ is around twice as much as the spectral efficiency of \emph{exhaustive search}.
It is due to $f_{comm}\approx 0.5$ for \emph{exhaustive search}, but $f_{comm}$ of \textbf{TC} is close to $1$ because of the small $T_{train}$. 
The spectral efficiency of \textbf{TC} at $(K_{op},N_{tr})=(20\%,5)$ outperforms the type B fingerprinting method at $(K_{op},N_{tr})=(40\%,10)$ by $0.7$ bit/s/Hz since our method provides more accurate beam prediction with even fewer known positions and fewer trained beams.
If we increase $K_{op}$ or $N_{tr}$, the improvement of \textbf{TC} is minor since $P_{pl}(\mathcal{S})$ is fairly low $(<10\%)$ in this region.


\vspace{-3.2mm}
\section{Conclusion}
In this paper, we propose a learning-based beam recommendation algorithm to reduce the training overhead for the position-aided beam alignment protocol.
We consider a scenario where the UE is located in an arbitrary position in which prior measurements may not be available in the database.
We propose a two-stage tensor completion to predict the received power, and then provide the set of recommended beams by ranking the predicted power of beams.
The two-stage tensor completion exploits both the low-rank and smoothness of the data.
The numerical results demonstrate that the proposed beam recommendation algorithm does improve the performance of beam-alignment over the state-of-the-art, by reducing the beam-training overhead.

\vspace{-4mm}
\bibliographystyle{IEEEtran}
\bibliography{IEEEabrv,reference} 
\end{document}